 \documentclass[final,5p,times,twocolumn]{elsarticle}  
 
 \usepackage{siunitx}
 \usepackage{upgreek}
\usepackage{amssymb}

\journal{Nuclear Instruments and Methods in Physics Research A}

\begin{document}

\begin{frontmatter}



\title{Development and operation of the CGEM Inner Tracker for the BESIII experiment}


\author[a,b,a]{Alberto Bortone}
    \affiliation[a]{organization={Turin University, department of Physics},
    addressline={Via Giuria 1}, 
    city={Turin},
    state={Italy}}
    \affiliation[b]{organization={Istituto Nazionale di Fisica Nucleare, Turin section},
    addressline={Via Giuria 1}, 
    city={Turin},
    state={Italy}}
\author[]{ on behalf of CGEM-IT Working Group}

\begin{abstract}
The extension of data acquisition for the BEijing Spectrometer (BESIII) experiment until at least 2030 has resulted in upgrades to both the accelerator and the detector.
An innovative Cylindrical Gas Electron Multiplier (CGEM) is under construction to upgrade the inner tracker, which is suffering from aging. The CGEM Inner Tracker (CGEM -IT) was designed to restore efficiency and enhance the reconstruction of the secondary vertexes position. Reconstruction in the magnetic field of 1 T requires an analog readout and an electronic contribution to the time resolution better than 5 ns. The entire system consists of about 10,000 electronic channels and must maintain a peak rate of 14 kHz/strip of signal hits for the innermost layer of the CGEM-IT. The CGEM readout system is based on the innovative TIGER ASIC, which is manufactured using 110 nm CMOS technology. A special readout chain consisting of GEM Read Out Cards (GEMROC) was developed for data acquisition.
Two of three layers, equipped with the final electronics, have been operating in Beijing since January 2020 collecting cosmic ray data in a test facility, remotely controlled by the Italian group due to the pandemic situation.
Since it was not possible to perform further tests on the cylindrical detector, in July 2021, a test beam was conducted at CERN with the final electronics configuration on a small prototype consisting of four triple-GEM planar detectors. 
In this presentation, the general status of the project CGEM-IT will be presented, with particular emphasis on the results of the test beam data collection.
\end{abstract}

\end{frontmatter}
\section{Introduction}
\label{sec:intro}
The BEijing Spectrometer (BESIII) is an experimental facility for  high-energy physics  at the Beijing Electron Positron Collider (BEPCII), with an extensive physics program that includes charmonium and charmonium-like states, charmed mesons and baryons, light hadron spectroscopy, $\tau$ physics, QCD and CKM parameters, baryon form factors, and new physics by studying rare and forbidden decays \citep{White_paper}. \\
BEPCII is a two-ring $e^+$ $e^-$ collider operating in the $\tau$-charm region with a center of mass energy between 2 and \SI{4.95}{\giga \electronvolt} and a luminosity of up to \SI{e33}{\cm^{-2}  \second ^{-1}}\citep{bepcII2009}.\\
The cylindrical core of the BESIII detector consists of multilayer drift chamber (MDC) \citep{Ablikim2010}, as a charged particle tracker. This detector has been taking data since 2009, but due to the high luminosity of the experiment, its performance is degrading, with a gain loss per year of about \SI{4}{\percent} for the innermost layers, despite the countermeasures take \citep{Dong_2016}. \\
Since BESIII is expected to be in operation for the next ten years, an upgrade is needed. An innovative solution to replace the aging inner MDC is to use a CGEM Inner Tracker (CGEM-IT).
\section{The CGEM-IT detector}
The gas electron multiplier (GEM) is a well developed micro pattern gas detector technology, with a wide range of applications in high-energy physics \citep{Sauli2016}.\\
In a triple-GEM detector, the multiplication structure takes place in a cascade of three GEM foils, allowing a higher gain while keeping the discharge risk low \cite{Bachmann:2002}.
\begin{figure}[h!]
\centering
\includegraphics[width=0.85\linewidth]{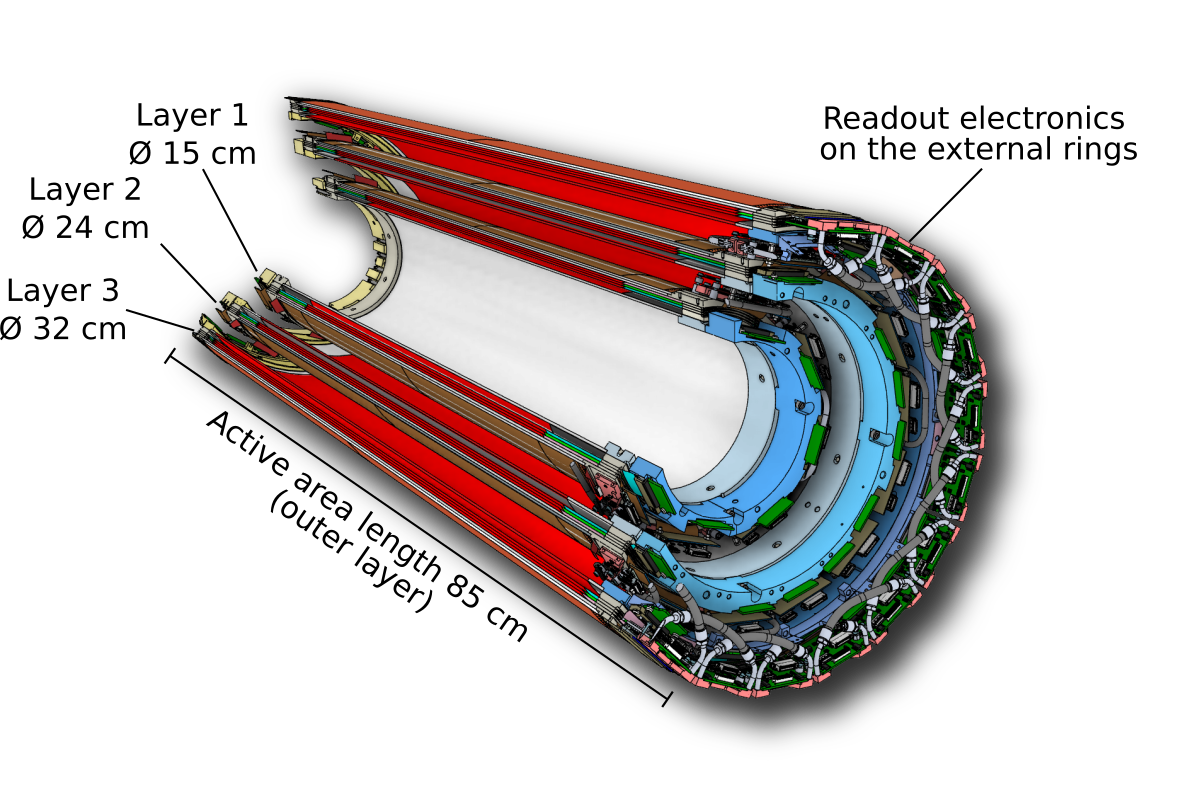}
  \caption{The CGEM-IT design}
  \label{fig:cgem}

\end{figure} 
The CGEM-IT consists of three coaxial layers of triple GEM (figure \ref{fig:cgem}). Each cylindrical detector layer is independently assembled,  has an autonomous gas enclosure, and can be operated in stand-alone mode.\\
The position reconstruction benefits from the combination of two different algorithms: the Charge Centroid extracts the average position by weighting the signal amplitude of the firing strips. The µ-TPC (micro Time-Projection Chamber) uses the drift gap like a time-projection chamber. The positions of the primary ionizations in the drift gap are reconstructed by knowing the drift velocity and the arrival time of the signal at the anode \cite{Riccardo2016}.  
The detector aims to achieve an r/$\Upphi$ resolution of about \SI{130}{\micro \meter}, a resolution along the beam direction of about \SI{350}{\micro \meter}, with a material budget around 1.5\% $X_0$ \cite{marcello2018,review_cgem}.
\section{The CGEM-IT readout}
The generated signal is induced on two arrays of strips to obtain a 2D readout: \SI{570}{\micro \meter} wide $\Upphi$-strips, parallel to the detector axis, and \SI{130}{\micro \meter} wide V strips that have a stereo angle with respect to the $\Upphi$ strips. The stereo angle is maximized according to the active area of each layer (between 31.0° to 46.7°), while the pitch, \SI{650}{\micro \meter}, is the same for both arrays and all the layers.\\
The reconstruction algorithms, the foreseen performance and the BESIII environment place special demands on the readout functions in terms of charge measurement, time resolution, and sustainable rate, requiring the development of a dedicated readout chain \cite{Bortone_2021} .\\
The signal information is acquired by the front-end ASIC TIGER (Torino Integrated GEM Electronics for Readout). The TIGER chips are mounted in pairs on Front-End Boards (FEB) and installed on the detector. Data and ASIC Low Voltage (LV) are fed through Data Low Voltage Patch Cards (DLVPC) by the GEM Read Out Cards (GEMROC).\\
The GEMROC boards receive  signals from the BESIII timing and trigger interface, communicate with the BESIII slow control via the Ethernet interface and via optical fibers with the GEM-Data Concentrators (GEM-DC). These cards build the events and communicate with the VME-based BESIII DAQ. The GEMROC boards also manage the  power supply and configuration of the FEBs, as well as the TIGER output data acquisition.
The GUFI  (Graphical Frontend User Interface) software has been developed to control the electronics and manage the data acquisition, while the CIVETTA  (Complete Interactive VErsatile Test Tool Analysis) software is used for on-run and off-line data validation and analysis \citep{Bortone_tesi}.
\section{Beam test}
\begin{figure}[h!]
\centering
\includegraphics[width=1\linewidth]{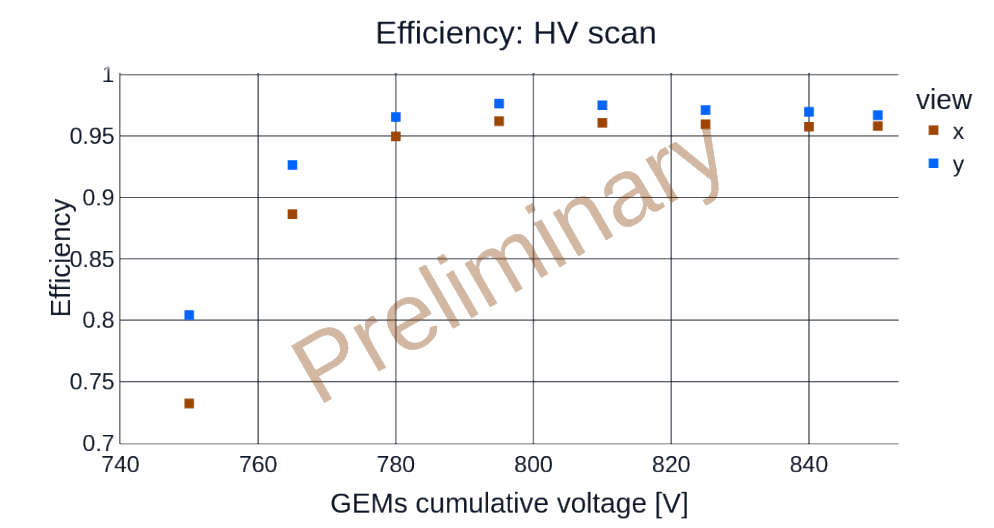}
  \caption{Detector efficiency as a function of the cumulative GEM voltage.}
  \label{fig:plot}

\end{figure}

A beam test was performed with planar triple-GEM detectors to investigate the system integration and evaluate the performance of the triple GEM detector equipped with the final TIGER + GEMROC readout chain. \\
The experiment was performed at the SPS North Area test beam facility at CERN, on the H4 beam line, with a 150 GeV $\pi$ beam and a \SI{80}{\giga \electronvolt} $\mu$ beam.\\ About 250M triggers have been acquired with beam incidence angles ranging from 0° to 45°, testing various detector and electronics settings.\\
The tested hodoscope consisted of four planar GEM detectors and two plastic scintillators coupled to PMTs in order to provide the trigger. The system was thoroughly tested by scanning different beam particle incidence angles, HV  and readout electronics settings. Figure \ref{fig:plot} shows the preliminary evaluation of the efficiency as a function of the GEM cumulative voltage. The efficiency increases with GEM voltage until it reaches a plateau. In analyzing these data, we  discovered a GEMROC firmware issue that can reduce the efficiency. A fix has been implemented and is currently being tested. 

\section*{Acknowledgment}
We acknowledge the support of the BESIIICGEM Project (645664) - H2020-MSCA-RISE-2014,
and of the FEST Project (872901) - H2020-MSCA-RISE-2019, which included INFN and IHEP as well as institutes from Mainz and Uppsala Universities.\\


\bibliographystyle{elsarticle-num-names} 
\bibliography{biblio}





\end{document}